\documentclass[12pt]{article}
\usepackage{amssymb}
\usepackage{amsfonts}
\usepackage{amsmath}
\usepackage{latexsym}
\usepackage[T1]{fontenc}
\usepackage[utf8]{inputenc}
\begin{document}

\title{On some states minimizing uncertainty relations:
		 A new look at these relations.}

\author{
K. Urbanowski\\
Institute of Physics,
University of Zielona G\'{o}ra,  \\
ul. Prof. Z. Szafrana 4a,
65-516 Zielona G\'{o}ra,
Poland\\
\hfill\\
e--mail:  K.Urbanowski@if.uz.zgora.pl, $\;$ k.a.urbanowski@gmail.com}
\maketitle


\begin{abstract}
The  Heisenberg--Robertson (HR) and Robertson--Schr\"{o}dinger (RS) uncertainty relations are studied.
Analyzing the RS uncertainty relation we found, that there can exist  a large set of states of the quantum system
under considerations, for which the lower bound of the product of the standard deviations of a pair of non--commuting
observables, $A$  and $B$,  is zero, and which differ from those described in the literature for the
HR uncertainty relation.  These states are not eigenstates of either the observable  $A$ or $B$.
The correlation function for these observables in such states is equal to zero.
We have also shown that the so--called "sum uncertainty relations" also do not provide any information
about lower bounds on the  standard deviations calculated for these states.
We additionally show that the uncertainty principle in its most general form has two faces: one is
that it is a lower bound on the product of standard deviations, and the other is that the product of
standard deviations is an upper bound on the modulus of the correlation function of a pair of the
non--commuting observables in the state under consideration.
\end{abstract}

\section{Introduction}
The Heisenberg uncertainty relations \cite{H,H2} was one of milestones
in understanding and interpreting
the quantum world.
In the  general, widely accepted case, the quantum uncertainty principle is understood as a lower bound   on the product of the standard deviations  $\Delta_{\phi} A$ and  $\Delta_{\phi} B$ of two non--commuting observables $A$ and $B$ calculated for a  given state, say $|\phi\rangle$,
\begin{equation}
\Delta_{\phi} A \,\cdot\, \Delta_{\phi} B \geq c> 0. \label{ur-c}
\end{equation}
Such an interpretation
follows from the derivation of the uncertainty relation made by Robertson
\cite{Robertson} and Schr\"{o}dinger \cite{Schrod-1930},  (see also \cite{M,Teschl}).   In a general case for an observable $F$ the standard deviation is defined as follows
\begin{equation}
\Delta_{\phi} F = \| \delta_{\phi} F|\phi\rangle\| \geq 0, \label{dF}
\end{equation}
where
$\delta_{\phi} F = F - \langle F\rangle_{\phi}\,\mathbb{I} $, and $\langle F\rangle_{\phi} \stackrel{\rm def}{=} \langle \phi|F|\phi\rangle$ is the average (or expected) value of an observable $F$ in a system whose state is represented by the normalized vector $|\phi\rangle \in {\cal H}$, provided that $|\langle\phi|F|\phi \rangle |< \infty$.
Equivalently:  $\Delta_{\phi} F \equiv \sqrt{\langle F^{2}\rangle_{\phi} - \langle F\rangle_{\phi}^{2}}$.
 The observable $F$ is represented by hermitian operator $F$ acting in a Hilbert space ${\cal H}$ of states $|\phi\rangle$. In general, the relation (\ref{ur-c}) results from basic assumptions of the quantum theory and from the geometry of Hilbert space \cite{M,Teschl}.

 Probably the most common form of inequality (\ref{ur-c}) is
 \begin{equation}
\Delta_{\phi} A \cdot \Delta_{\phi} B\;\geq\;\frac{1}{2} \left|\langle [A,B] \rangle_{\phi} \right|,\label{R1}
\end{equation}
which holds for any two observables, $A$ and $B$, represented by non--commuting hermitian operators $A$ and $B$ acting in ${\cal H}$ (see \cite{Robertson} and also \cite{Schrod-1930,M,Teschl}), such that $[A,B]$ exists and $|\phi\rangle \in {\cal D}(AB) \bigcap {\cal D}(BA)$, (${\cal D}({\cal O})$ denotes the domain of an operator $\cal O$ or of a product of operators).
For the convenience of readers, we will recall its derivation.
The derivation the rigorous one. Indeed, the first step is to use  the Cauchy--Schwarz inequality
\begin{equation}
\left\|\;|\psi_{A}\rangle \right\| \,\left\|\;|\psi_{B}\rangle\right\| \geq \left| \langle \psi_{A}|\psi_{B} \rangle \right|, \label{S}
\end{equation}
and write it appropriately for vectors $|\psi_{A}\rangle = \delta_{\phi} A |\phi\rangle $ and $|\psi_{B}\rangle = \delta_{\phi} B|\phi\rangle $  (see, e.g. \cite{Teschl}):
\begin{equation}
\left\| \delta_{\phi} A |\phi\rangle \right\|^{2}\;\left\| \delta_{\phi} B|\phi\rangle \right\|^{2}\,\geq \,\left|\langle\phi| \delta_{\phi} A\;\delta_{\phi} B|\phi \rangle \right|^{2}, \label{dAdB}
\end{equation}
which holds  for all $|\phi\rangle \in {\cal D}(AB) \bigcap {\cal D}(BA)$. (The equality in (\ref{S}), or in (\ref{dAdB}), holds when $|\psi_{A}\rangle = z |\psi_{B}\rangle$, $z \in \mathbb{C}$).
The next step is to transform the right side of Eq. (\ref{dAdB}) as follows:
\begin{eqnarray}
\left|\langle\phi| \delta_{\phi} A\;\delta_{\phi} B|\phi \rangle \right|^{2} & = & \left[ \Re\,(\langle \phi|\delta_{\phi} A\;\delta_{\phi} B|\phi \rangle) \right]^{2} +
\left[ \Im\,(\langle \phi|\delta_{\phi} A\;\delta_{\phi} B|\phi \rangle) \right]^{2}, \label{R1a} \\
&=& \frac{1}{4} \left(\langle\phi|( \delta_{\phi} A\;\delta_{\phi} B\,+\, \delta_{\phi} B\;\delta_{\phi} A)|\phi \rangle \right)^{2} \nonumber \\
&  &\;\;\;\; \;\;\;+\; \frac{1}{4} \left|\langle\phi|( \delta_{\phi} A\;\delta_{\phi} B\,-\, \delta_{\phi} B\;\delta_{\phi} A)|\phi \rangle \right|^{2} \nonumber \\
 &\equiv &
 \frac{1}{4} \left(\langle\phi|( \delta_{\phi} A\;\delta_{\phi} B\,+\, \delta_{\phi} B\;\delta_{\phi} A)|\phi \rangle \right)^{2} \nonumber\\
 & & \;\;\;\; \;\;\;+\;
 \frac{1}{4} \left|\langle\phi|[A,B]|\phi \rangle \right|^{2}  \label{Sch-1}\\
 & \geq & \frac{1}{4} \left|\langle\phi|[A,B]|\phi \rangle \right|^{2}, \label{R1+1}
\end{eqnarray}
where $\Re\,(z)$ denotes the real part of the complex number $z$ and $\Im\,(z)$ is the imaginary part of $z$.
The property $[\delta_{\phi} A, \delta_{\phi} B] = [A,B]$  taking place  for all $|\phi\rangle \in {\cal D}(AB) \bigcap {\cal D}(BA)$ was used in (\ref{Sch-1}).

Note that  if to  use definition (\ref{dF}) and then replace the right hand side of Eq. (\ref{dAdB}) by (\ref{Sch-1}) then one obtains the uncertainty relation of the type derived by Schr\"{o}dinger \cite{Schrod-1930}:
\begin{equation}
 (\Delta_{\phi} A )^{2}\, \cdot\,( \Delta_{\phi} B)^{2}\,\geq
\frac{1}{4} \left(\langle\phi|( \delta_{\phi} A\;\delta_{\phi} B\,+\, \delta_{\phi} B\;\delta_{\phi} A)|\phi \rangle \right)^{2} \,+\,
 \frac{1}{4} \left|\langle\phi|[A,B]|\phi \rangle \right|^{2},  \label{Sch-2}
\end{equation}
or,   equivalently, in more familiar form,
\begin{equation}
 (\Delta_{\phi} A )^{2}\, \cdot\,( \Delta_{\phi} B)^{2}\,\geq
 \left(\frac{ \langle(AB +BA)\rangle_{\phi}}{2} -  \langle A\rangle_{\phi}\,\langle B \rangle_{\phi}\right)^{2} \,+\,
  \left|\frac{\langle [A,B] \rangle_{\phi}}{2} \right|^{2}.  \label{Sch-3}
\end{equation}
As it can be seen relations (\ref{Sch-2}), (\ref{Sch-3})  seem to be  more general and precise
than the relation (\ref{R1}). It is because the right hand sides of (\ref{Sch-2}), (\ref{Sch-3}) are strictly equivalent to the right hand side of Eq. (\ref{dAdB}).

Now if  one replaces the right hand side in Eq. (\ref{dAdB}) by (\ref{R1+1}) then one obtains the uncertainty relation (\ref{R1})  as a result.
The transformations leading to formulas (\ref{R1+1}), (\ref{Sch-1}) prove that the Heisenberg--Robertson (HR) uncertainty relation  (\ref{R1}) is less precise than the Schr\"{o}dinger uncertainty relations  (\ref{Sch-2}), (\ref{Sch-3}). The uncertainty relations (\ref{R1}) and  (\ref{Sch-2}), (\ref{Sch-3}) are state--depended and
only these types of relations will be considered in the following parts of the paper.

\section{Analysis and results}

Let us analyze now  properties of inequalities (\ref{R1}) and (\ref{dAdB}) --- (\ref{Sch-3}).
Firstly,
as is well known, the right side of  (\ref{R1}) becomes zero when $[A,B] \neq 0$ and vector $|\phi\rangle$ is an eigenvector of $A$ or $B$. If to
assume that, e.g.,  $|\phi \rangle = |\psi_{b}\rangle$ is a normalized eigenvector of  $B$ for the eigenvalue $b$, that is that $B|\psi_{b}\rangle = b |\psi_{b}\rangle$, then one immediately finds that $\langle \psi_{b}|B|\psi_{b} \rangle = b$, and $\langle \psi_{b}|AB|\psi_{b}\rangle = b \langle \psi_{b}|A|\psi_{b}\rangle$ and also
 $\delta_{\psi_{b}} B|\psi_{b}\rangle \equiv 0$.
As a result we have that
 $\langle\psi_{b}|[A,B]|\psi_{b}\rangle \equiv 0$ and that
 $\Delta_{\psi_{b}}(B) = 0$. The same effect, i. e. that the right hand  side of (\ref{R1}) is equal to zero for $|\phi\rangle$ being an eigenvector of $A$ or of $B$,
  takes place in the case of more general inequalities  (\ref{Sch-2}), (\ref{Sch-3}). It is because, the two sides of the Cauchy--Schwarz inequality (\ref{S}) are equal to zero if $|\psi_{A}\rangle = 0$ (or $|\psi_{B}\rangle =0$), and in the considered case $|\phi\rangle = |\psi_{b}\rangle$ there is $|\psi_{B} \rangle \equiv \delta_{\psi_{b}} B|\psi_{b}\rangle \equiv 0$ in (\ref{dAdB}).

The second,
   much more interesting, non--trivial case of the vanishing  right side of the Cauchy--Schwarz inequality (\ref{S}), and thus the inequality (\ref{dAdB}),  is when  $[A,B] \neq0$ and vectors both appearing therein  $|\psi_{A}\rangle = \delta_{\phi} A\,|\phi\rangle, |\psi_{B}\rangle = \delta_{\phi} B\,|\phi\rangle$ are nonzero and $\delta_{\phi} A\,|\phi\rangle \,\perp\,\delta_{\phi} B\,|\phi\rangle$.
  This implies that there are
    $\Delta_{\phi}A > 0, \, \Delta_{\phi}B > 0$ and $\Delta_{\phi}A \,\cdot\,\Delta_{\phi}B \geq |\langle \phi| \delta_{\phi} A\,\delta_{\phi} B|\phi\rangle| \equiv 0$
    instead of inequalities (\ref{ur-c}),  (\ref{R1}) and  (\ref{Sch-2}), (\ref{Sch-3}).
  So in this case there is no a lower bound for  product of standard deviations  $\Delta_{\phi}A $ and $\Delta_{\phi}B$.
  More precisely: it is equal to zero.
  The number of  such $|\psi_{A}\rangle = \delta_{\phi} A\,|\phi\rangle, |\psi_{B}\rangle = \delta_{\phi} B\,|\phi\rangle$, strictly speaking the number of corresponding vectors $|\phi\rangle$, can be extremely  large depending on the dimension of the state space. As it can be easily seen there must be: $\langle\phi|\psi_{A}\rangle = \langle \phi| \delta_{\phi} A\,|\phi\rangle \equiv 0$ and similarly  $\langle\phi|\psi_{B}\rangle = \langle \phi| \delta_{\phi} B\,|\phi\rangle \equiv 0$. In other words there are: $|\psi_{A}\rangle = \delta_{\phi} A\,|\phi\rangle \stackrel{\rm def}{=} \alpha_{A}|\phi_{A}^{\perp}\rangle \,\perp \,|\phi\rangle$ and $|\psi_{B}\rangle = \delta_{\phi} B\,|\phi\rangle \stackrel{\rm def}{=}\alpha_{B}|\phi_{B}^{\perp}\rangle \,\perp \,|\phi\rangle$, where $\|\,|\phi_{A}^{\perp}\rangle\| = \|\,|\phi_{B}^{\perp}\rangle\| = 1$, (see also \cite{Aha}).
  It can be easily shown that $\alpha_{A} = \Delta_{\phi}A$ and  $\alpha_{B} =  \Delta_{\phi}B$, which means that
  \begin{equation}
  \delta_{\phi}A|\phi\rangle \equiv \Delta_{\phi}A\,|\phi_{A}^{\perp}\rangle, \;\;{\rm and}\;\; \delta_{\phi}B|\phi\rangle \equiv \Delta_{\phi}B\,|\phi_{B}^{\perp}\rangle. \label{DA+DB}
  \end{equation}
  As we have already seen, we analyze the condition  $\langle \psi_{A}|\psi_{B}\rangle  \equiv \langle \phi_{A}^{\perp}|\phi_{B}^{\perp}\rangle = 0$, that is $|\psi_{A}\rangle \equiv \alpha_{A}|\phi_{A}^{\perp}\rangle \,\perp\,|\psi_{B}\rangle  \equiv \alpha_{B}|\phi_{B}^{\perp}\rangle$. Summing up: For a given $|\phi\rangle$ we have $|\phi\rangle \perp |\phi_{A}^{\perp}\rangle,\; |\phi\rangle \perp |\phi_{B}^{\perp}\rangle$ and $ |\phi_{A}^{\perp}\rangle \perp |\phi_{B}^{\perp}\rangle$. This problem has no non--zero solutions in two--dimensional space.  They exist when the dimension of the state space is not less than 3.

States $|\psi_{A}\rangle = \delta_{\phi} A\,|\phi\rangle, |\psi_{B}\rangle = \delta_{\phi} B\,|\phi\rangle$
 have other interesting properties.
 Analyzing the left side of equation (\ref{R1a}) we find
\begin{equation}
\langle \phi| \delta_{\phi} A\,\delta_{\phi} B|\phi\rangle \equiv \langle AB \rangle_{\phi} - \langle A\rangle_{\phi}\,\langle B\rangle_{\phi} \equiv {\cal C}_{\phi}(A,B), \label{co1}
\end{equation}
where $\langle AB \rangle_{\phi} = \langle \phi|AB|\phi\rangle $ and ${\cal C}_{\phi}(A,B)$ is a quantum version of the correlation function known also as a covariance. Here defining the correlation function ${\cal C}_{\phi}(A,B)$ we follow, e. g.  \cite{Poz,Khr} and others.

In fact, the  correlation function in the large literature is defined as the  matrix element $\langle \phi| \delta_{\phi} A\,\delta_{\phi} B|\phi\rangle$ (see, eg. \cite{Bei,Rei,Roh}).
Simply,
there is:   $\langle \phi| \delta_{\phi} A\,\delta_{\phi} B|\phi\rangle = \langle \phi|\left[\left( A - \langle A\rangle_{\phi} \right)  \left( B - \langle B\rangle_{\phi}\right)\right]|\phi\rangle \equiv {\cal C}_{\phi}(A,B)$.
The identity (\ref{co1}) means that
\begin{equation}
\langle \phi| \delta_{\phi} A\,\delta_{\phi} B|\phi\rangle  = 0 \;\Leftrightarrow\; \langle \phi |AB| \phi\rangle - \langle A\rangle_{\phi}\,\langle B\rangle_{\phi} = 0. \label{co2}
\end{equation}
Now,
if we use definitions (\ref{dF}) in (\ref{dAdB}) and (\ref{co2}) then we get the following inequality
that coincides with Schr\"{o}dinger's uncertainty relations (\ref{Sch-2}) and (\ref{Sch-3}) and which is another version of it (see also \cite{Mac1}):
\begin{equation}
\Delta_{\phi} A \cdot \Delta_{\phi} B\;\geq\;\left|{\cal C}_{\phi}(A,B) \right|. \label{gen-ur}
\end{equation}
Note that this inequality shows a direct connection between
the product of standard deviations
and the correlation function ${\cal C}_{\phi}(A,B)$.
Equivalently,
\begin{equation}
\Delta_{\phi} A \cdot \Delta_{\phi} B\;\geq\;\left|\langle AB \rangle_{\phi} - \langle A\rangle_{\phi}\,\langle B\rangle_{\phi} \right|. \label{gen-ur1a}
\end{equation}

If the system is in a state $|\phi\rangle$ such that $\Delta_{\phi}A > 0$ and $\Delta_{\phi}B >0$, then the inequality (\ref{gen-ur}) can be written in an equivalent, useful in some applications,  form as
\begin{equation}
\mathfrak{r}_{\phi}(A,B)  \stackrel{\rm def}{=}   \frac{\left|{\cal C}_{\phi}(A,B) \right|}{\Delta_{\phi} A \cdot \Delta_{\phi} B} \leq 1, \label{r1}
\end{equation}
which looks like  a variant  of  Pearson's  coefficient, i.e the correlation coefficient (see, e.g. \cite{Rei,Mac1,Jeb}) or rather its quantum modification.
If we take into account equations (\ref{DA+DB}), $ {\cal C}_{\phi}(A,B)$ can be written as
\begin{equation}
{\cal C}_{\phi}(A,B) = (\Delta_{\phi}A) \,(\Delta_{\phi }B) \,\langle \phi_{A}^{\perp}| \phi_{B}^{\perp}\rangle, \label{co3}
\end{equation}
 and thus the coefficient $\mathfrak{r}_{\phi}(A,B) $ is simply
\begin{equation}
\mathfrak{r}_{\phi}(A,B) \equiv |\langle \phi_{A}^{\perp}| \phi_{B}^{\perp}\rangle|. \label{r01}
\end{equation}
So, for a given $A$ and $B$,  the coefficient $\mathfrak{r}_{\phi}(A,B) $  describes the intensity of transitions between states $|\phi_{A}^{\perp}\rangle$ and $|\phi_{B}^{\perp}\rangle$ orthogonal  to $|\phi\rangle$,
and
\begin{equation}
\left(\mathfrak{r}_{\phi}(A,B)\right)^{2} = |\langle \phi_{A}^{\perp}| \phi_{B}^{\perp}\rangle|^{2}, \label{p1}
\end{equation}
is the transition probability  from state $|\phi_{A}^{\perp}\rangle$ to  $|\phi_{B}^{\perp}\rangle$  (and vice versa).
Thus equations (\ref{r01}) and (\ref{p1}) tell us what the Pearson coefficient (its quantum version) physically means in the quantum world.

In some papers the covariance is defined as the real part of ${\cal C}_{\phi}(A,B)$. That is as:  ${\rm cov}_{\phi}(A,B) = \Re\,[{\cal C}_{\phi}(A,B)]$ (see, e. g.,  \cite{Mac1,Deb}).
It should be noted that the function ${\rm cov}_{\phi}(A,B)$ is just the classical part of the quantum version of the covariance ${\cal C}_{\phi}(A,B) $ and does not describe all properties of quantum systems.
Indeed, from equations (\ref{r1}), (\ref{co3}), (\ref{r01})
and (\ref{p1}) we get that
\begin{equation}
|\langle \phi_{A}^{\perp}| \phi_{B}^{\perp}\rangle|^{2}
\equiv \left(\frac{{\rm cov}_{\phi}(A,B)}{\Delta_{\phi} A \cdot \Delta_{\phi} B}\right)^{2} + \left(\frac{\Im\,[{\cal C}_{\phi}(A,B)]}{\Delta_{\phi} A \cdot \Delta_{\phi} B}\right)^{2}. \label{p2}
\end{equation}
This result shows that if we limit ourselves only to the function ${\rm cov}_{\phi}(A,B)$ when studying correlations in quantum systems, we lose a part of the information about the correlation between observables $A$ and $B$ in the state $|\phi\rangle$, which is hidden in the term  containing $\Im\,[{\cal C}_{\phi}(A,B)] $ in equation (\ref{p2}),  which may lead to  wrong conclusions.
Simply,
it may happen that the system will be in such state $|\phi\rangle$ that ${\rm cov}_{\phi}(A,B) = 0$ but $\Im\,[{\cal C}_{\phi}(A,B)] \neq 0 $, which will cause that $|\langle \phi_{A}^{\perp}| \phi_{B}^{\perp}\rangle|^{2} \neq 0$
and therefore that $\left(\mathfrak{r}_{\phi}(A,B)\right)^{2} \neq 0$.

It is easy to find some formal properties of ${\cal C}_{\phi}(A,B) $ and of $  \mathfrak{r}_{\phi}(A,B)$:
\begin{eqnarray}
{\cal C}_{\phi}(A,A)& =& (\Delta_{\phi}A)^{2}, \nonumber \\
{\cal C}_{\phi}(A,B)& =& [{\cal C}_{\phi}(B,A)]^{\ast}, \nonumber \\
{\cal C}_{\phi}(A,B_{1} +B_{2}) &=& {\cal C}_{\phi}(A,B_{1})+ {\cal C}_{\phi}(A,B_{2}),\label{cov3}
\end{eqnarray}
\begin{equation}
\mathfrak{r}_{\phi}(A,A) =1, \;\;\mathfrak{r}_{\phi}(A,B) = \mathfrak{r}_{\phi}(B,A). \label{r-2}
\end{equation}
It should be noted that $\mathfrak{r}_{\phi}(A,A) = 1 $ even though $[A,A]$=0, which means that the right side of the HR relation (\ref{R1}) is equal to zero.
Moreover, $\mathfrak{r}_{\phi}(A,B) = 0 $ if only there exists such a state $|\phi\rangle$ that $\delta_{\phi} A|\phi\rangle \perp \delta_{\phi} B|\phi\rangle$ despite the fact that $[A,B] \neq 0$.
The result $\mathfrak{r}_{\phi}(A,B) = 0 $ means that  observables $ A$ and $B$ are uncorrelated in the state $|\phi\rangle$. However, this does not exclude that in a state $|\psi\rangle \neq |\phi\rangle$ there may be $\mathfrak{r}_{\psi}(A,B) > 0$ and then the value $\mathfrak{r}_{\psi}(A,B)$ describes the level of a correlation.
The case $\mathfrak{r}_{\phi}(A,B) =1 $ occurs when   the equality holds in relations (\ref{dAdB}), (\ref{Sch-2}), (\ref{Sch-3}) and (\ref{gen-ur}),  i. e. if
 \begin{equation}
 \delta_{\phi}A|\phi\rangle = z \delta_{\phi}B|\phi\rangle. \label{i1}
 \end{equation}
  (State vectors that are solutions to this equation are called intelligent states (see, e.g., \cite{Jac,fic1,Sh})).
 Indeed,   for this type of states, in inequality (\ref{gen-ur}) we have the equality: $\Delta_{\phi}A \cdot \Delta{\phi}B = |{\cal C}_{\phi}(A,B))|$, which in turn means that for these states the Pearson's coefficient   is equal to $1$: $\mathfrak{r}_{\phi}(A,B) =1$.
 So, the conclusion is
 that the non-commuting observables $A$ and $B$ are fully correlated  in the state  $|\phi\rangle$ solving the Eq. (\ref{i1}), i.e., in the intelligent states of the system.

Note that using definition (\ref{dF})  and the Eq. (\ref{i1}) one can also immediately find that
\begin{equation}
 |z| = \frac{\Delta_{\phi}A}{\Delta_{\phi}B}. \label{|z|}
\end{equation}
This leads to the conclusion that for a certain class of intelligent states for which $\Delta_{\phi}A = \Delta_{\phi}B$, the property $|z| = 1$ holds.

Another observation concerns Eq. (\ref{R1+1}).
Suppose  that $[A,B] \neq 0$ and we found such a state $|\phi\rangle = |\phi_{0}\rangle$, which is not an eigenstate of $A$ or of $B$, that
$\langle \phi_{0}|[A,B]|\phi_{0} \rangle = 0$.
It appears that
this property of the state $|\phi_{0}\rangle$ need not imply that
$ \delta_{\phi_{0}} A|\phi_{0}\rangle \perp  \delta_{\phi_{0}} B|\phi_{0}\rangle$, i. e. that
$\langle \phi_{0}| \delta_{\phi_{0}} A\,\delta_{\phi_{0}} B|\phi_{0}\rangle =0  $.  Indeed, from equations (\ref{R1a}) --- (\ref{R1+1}) it follows that the result $\langle \phi_{0}|[A,B]|\phi_{0} \rangle = 0$ is equivalent
 to the condition $\Im(\langle \phi_{0}| \delta_{\phi_{0}} A\;\delta_{\phi_{0}} B|\phi_{0}\rangle) = 0$, but this does not mean that also $\Re(\langle \phi_{0}| \delta_{\phi_{0}} A\;\delta_{\phi_{0}} B|\phi_{0}\rangle) = 0$ and it may happen that $\Re(\langle \phi_{0}| \delta_{\phi_{0}} A\;\delta_{\phi_{0}} B|\phi_{0}\rangle) \neq 0$ which implies that $\langle \phi_{0}| \delta_{\phi_{0}} A\;\delta_{\phi_{0}} B|\phi_{0}\rangle \neq 0$, and the number of such cases does not have to be small.
On the other hand, from the same equations it follows that if $\langle \phi| \delta_{\phi} A\,\delta_{\phi} B|\phi\rangle = 0 $ then there must always be $ \langle \phi|[A,B]|\phi \rangle = 0$.

\section{Examples}

  As an illustration of the cases under consideration, let us analyze the following example:
  Assume that
  $A=\lambda_{3}$ and $B = \lambda_{4}$,
  where
  $\lambda_{3}, \lambda_{4}$ are Gell--Mann matrices:
\begin{equation}
\lambda_{3} = \left(
                \begin{array}{ccc}
                  1 & 0 & 0 \\
                  0 & -1 & 0 \\
                  0 & 0 & 0 \\
                \end{array}
              \right),\;\;
\lambda_{4} = \left(
                \begin{array}{ccc}
                  0 & 0 & 1 \\
                  0 & 0 & 0 \\
                  1 & 0 & 0 \\
                \end{array}
              \right),\;\;
\lambda_{5} =   \left(
                \begin{array}{ccc}
                  0 & 0 & i \\
                  0 & 0 & 0 \\
                  -i & 0 & 0 \\
                \end{array}
              \right).
\end{equation}
Here $\lambda_{5}$ is also Gell--Mann matrix.
They are self--adjoint and do not commute,
$
[\lambda_{3},\lambda_{4}]=-i\lambda_{5} \neq  0.
$
For these matrices the inequalities (\ref{dAdB}), (\ref{gen-ur}) takes the following form,
\begin{equation}
\left\|\delta_{\phi} \lambda_{3}\,|\phi\rangle \right\|^{2}\, \cdot \,\left\|\delta_{\phi} \lambda_{4}|\phi\rangle\right\|^{2} \geq \left| \langle \phi|\delta_{\phi}\lambda_{3}\,\delta_{\phi}\lambda_{4}|\phi \rangle \right|^{2} \equiv |{\cal C}_{\phi}(\lambda_{3}, \lambda_{4})|^{2}.
\label{dAdB-l}
\end{equation}
Let us choose for simplicity,
\begin{equation}
|\phi\rangle = |\phi_{1}\rangle =\frac{1}{N}\,\left(
                            \begin{array}{c}
                                 a \\
                                 b \\
                                 0 \\
                            \end{array}
                            \right), \label{phi-l}
\end{equation}
where $N^{2} = |a|^{2} + |b|^{2}$
and $a,b\in \mathbb{C}$. Using (\ref{phi-l}) one gets
\begin{equation}
\delta_{\phi_{1}} \lambda_{3}|\phi_{1}\rangle = \frac{1}{N^{3}} \,                    \left(
                                                         \begin{array}{c}
                                                          2|b|^{2} a \\
                                                          -2|a|^{2} b \\
                                                           0 \\
                                                         \end{array}
                                                       \right) \neq 0,
                                                        \label{d-l-3}
\end{equation}
and
\begin{equation}
\delta_{\phi_{1}} \lambda_{4}|\phi_{1}\rangle = \frac{1}{N} \,                    \left(
                                                         \begin{array}{c}
                                                          0 \\
                                                          0 \\
                                                           a \\
                                                         \end{array}
                                                       \right) \neq 0,
                                                        \label{d-l-4}
\end{equation}
(where $\delta_{\phi_{1}} \lambda_{k}|\phi_{1}\rangle = (\lambda_{k} - \langle \phi_{1}|\lambda_{k}|\phi_{1}\rangle)|\phi_{1}\rangle$ and $k = 3,4$),
which leads to the result ${\cal C}_{\phi_{1}}(\lambda_{3},\lambda_{4}) = \langle \phi_{1}|\delta_{\phi_{1}} \lambda_{3}\;\delta_{\phi_{1}} \lambda_{4}|\phi_{1} \rangle = 0$, and $(\Delta_{{\phi}_{1}}\lambda_{3})^{2} = { \left\| \delta_{\phi_{1}} \lambda_{3}|\phi_{1}\rangle\right\|}^{2} \neq 0$,
$(\Delta_{{\phi}_{1}}\lambda_{4})^{2} = { \left\|\delta_{\phi_{1}} \lambda_{4}|\phi_{1}\rangle\right\|}^{2} \neq 0$.
Hence one concludes that for $|\phi_{1}\rangle$ the inequality (\ref{dAdB-l}) and thus relations (\ref{R1}), (\ref{Sch-2}), (\ref{Sch-3})
take the following form,
\begin{equation}
\Delta_{\phi_{1}} \lambda_{3}\,\cdot\,\Delta_{\phi_{1}} \lambda_{4} \geq 0.\label{R2-la}
\end{equation}
Note that this result holds for any $a \neq 0$ and $b \neq 0$ defining the vector $|\phi_{1}\rangle$. This shows that the number of such vectors may be very large.

Now let's consider another example. Again, let $A = \lambda_{3}$ and $B = \lambda_{4}$.
For these matrices the inequality (\ref{R1}) takes the following form,
\begin{equation}
\Delta_{\phi} \lambda_{3}\,\cdot\,\Delta_{\phi} \lambda_{4} \geq \frac{1}{2}\,\left|\langle [\lambda_{3},\lambda_{4}] \rangle_{\phi} \right| \equiv
\frac{1}{2}\,\left|\langle \lambda_{5} \rangle_{\phi}\right|, \label{R1a-lambda}
\end{equation}
Assume that
\begin{equation}
|\phi \rangle \;\;=\;\; |\phi_{2}\rangle = \frac{1}{\sqrt{3}}\,\left(
                                                         \begin{array}{c}
                                                           1 \\
                                                           1 \\
                                                           1 \\
                                                         \end{array}
                                                       \right), \label{phi-2-lambda}
\end{equation}
which leads to the result $\left|\langle \lambda_{5} \rangle_{\phi_{2}} \right| = 0$, and hence one concludes that for $|\phi_{2}\rangle$ the inequality (\ref{R1a-lambda})                                                        takes the following form
\begin{equation}
\Delta_{\phi_{2}} \lambda_{3}\,\cdot\,\Delta_{\phi_{2}} \lambda_{4} \geq 0, \label{R2-lambda}
\end{equation}
where $\Delta_{\phi_{2}}\lambda_{3} \neq 0$ and  $\Delta_{\phi_{2}}\lambda_{4} \neq 0$.

Elementary calculations show that in this case
\begin{eqnarray}
\langle \phi_{2} |\delta_{\phi_{2}} A\;\delta_{\phi_{2}} B|\phi_{2}\rangle & =&  \langle \phi_{2} |\delta_{\phi_{2}} \lambda_{3}\;\delta_{\phi_{2}} \lambda_{4}|\phi_{2}\rangle\nonumber \\
&\equiv&   {\cal C}_{\phi_{2}}(\lambda_{3},\lambda_{4}) = \langle \phi_{2}|\lambda_{3}\lambda_{4}|\phi_{2}\rangle = \frac{1}{3}.
\label{dl-dl}
 \end{eqnarray}
(It is because $\langle \phi_{2}|\lambda_{3}|\phi_{2}\rangle = 0$). Finally we get
\begin{equation}
\Delta_{\phi_{2}} \lambda_{3}\,\cdot\,\Delta_{\phi_{2}} \lambda_{4}\; \geq |\langle \phi_{2} |\delta_{\phi} \lambda_{3}\;\delta_{\phi} \lambda_{4}|\phi_{2}\rangle|=  {\cal C}_{\phi_{2}}(\lambda_{3},\lambda_{4}) = \frac{1}{3} \neq 0. \label{dl-dl-2}
\end{equation}
Concluding: As can be seen from results  (\ref{R1a-lambda}) --- (\ref{dl-dl-2}), the existence of such a state $|\phi\rangle$ that $\langle[A,B]\rangle_{\phi} = 0$ does not necessarily guarantee that the right--hand side of the more general inequalities (\ref{Sch-2}), (\ref{Sch-3}), (\ref{gen-ur}) will also be equal to zero.

The example discussed in this Section is one of the mathematically simplest  examples and at the same
time it is a non--trivial illustration of the problems considered in the previous Section. On the other hand,
the Gell--Mann $\lambda$ matrices are the natural basis for the Hilbert space of
Hermitian operators acting on the states of a three--level systems (qutrits). States of the type (\ref{phi-l}), (\ref{phi-2-lambda}) and similar are considered when analyzing three--level systems (see, eg. \cite{Jaf,Kis,Bo,Jia,Ken} and other papers).

\section{Discussion}

Recently, uncertainty relations for sums of standard deviations or  variances have been studied in the literature (see, e. g. (see, e. g. \cite{Pat,Mac,Chi,Lei,Dad} and many other papers), which are considered to be
more useful
than the HR or Schr\"{o}dinger uncertainty relations. These studies are motivated by the fact that the latter do not provide any bounds on the standard deviations of either the observables $A$ or $B$ when
the system is in the state described by the eigenfunction of the observable $B$ (or $A$)
which causes that
the right-hand side of (\ref{R1}) inequality vanishes. Relations of this type
are called "sum uncertainty relations".
Let us now look at some of these "sum uncertainty relations" and see if they are sensitive to the effects described above.
Their mathematical basis is the triangle inequality in the Hilbert state space: There is
\begin{equation}
\left\|\,|\psi_{1}\rangle\right\| + \left\|\,|\psi_{2}\rangle\right\|\,\geq\,\left\| \,|\psi_{1}\rangle + |\psi_{2}\rangle \right\|, \label{n-t}
\end{equation}
for each pair of vectors $|\psi_{1}\rangle, |\psi_{2}\rangle \in \cal{H}$.   It is usually derived in Hilbert space using the Schwarz inequality (\ref{S}) ---  see, e. g.,  \cite{Teschl,Pat,Dad,Cer}.
If we now substitute $|\psi_{A}\rangle = \delta_{\phi} A |\phi\rangle $ and $|\psi_{B}\rangle = \delta_{\phi} B|\phi\rangle $
 into (\ref{n-t}) instead of $|\psi_{1}\rangle$ and  $|\psi_{2}\rangle$, we get the "sum uncertainty relation" derived in \cite{Pat},
 \begin{equation}
 \left\|\delta_{\phi} A |\phi\rangle\right\| + \left\|\delta_{\phi} B\phi\rangle\right\|\,\geq\,\left\| \delta_{\phi} A|\phi\rangle + \delta_{\phi} B |\phi\rangle \right\| \equiv \left\|\delta_{\phi}(A + B)|\phi\rangle \right\|,
 \label{n-t2}
\end{equation}
that is,
\begin{equation}
\Delta_{\phi}A + \Delta_{\phi}B \geq \Delta_{\phi}(A + B). \label{n-t3}
\end{equation}
When looking for inequalities of the "sum uncertainty relation" type, one can also use the triangle inequality of the second kind \cite{Mac,Hsu}:
\begin{equation}
\left\|\,|\psi_{1}\rangle\right\|^{2} + \left\|\,|\psi_{2}\rangle\right\|^{2}\,\geq\,\frac{1}{2}\left\| \,|\psi_{1}\rangle + |\psi_{2}\rangle \right\|^{2}, \label{n-2t}
\end{equation}
which leads to the following sum uncertainty relation,
\begin{equation}
(\Delta_{\phi}A)^{2} + (\Delta_{\phi}B)^{2} \geq  \frac{1}{2} \left[\,\Delta_{\phi}(A + B)\right]^{2}. \label{n-t4}
\end{equation}
This type of a generalization of the inequality (\ref{n-t3}) has been studied in many papers (see, e. g. \cite{Mac,Chi,Lei,Bin,Son} and many others).

Now, let $|\phi \rangle = |\psi_{b}\rangle$, where $|\psi_{b}\rangle $ is and eigenvector of the observable $B$ for the eigenvalue $b$. Then, as it has been shown earlier, $\delta_{\psi_{b}} B|\psi_{b}\rangle = 0$, and therefore $\Delta_{\psi_{b}}B =0$, and  $\delta_{\psi_{b}}(A+B)|\psi_{b}\rangle = \delta_{\psi_{b}} A|\psi_{b}\rangle + \delta_{\psi_{b}} B|\psi_{b}\rangle \equiv \delta_{\psi_{b}} A|\psi_{b}\rangle$, and as a result we have  $\Delta_{\psi_{b}}(A + B) \equiv \Delta_{\psi_{b}}A$.
This means that even though the right--hand side of the inequalities (\ref{n-t3}) and (\ref{n-t4}) is non--zero,
these inequalities become trivial: $\Delta_{\psi_{b}}A  \geq  \Delta_{\psi_{b}}A $ (or $(\Delta_{\psi_{b}}A)^{2}  \geq  \frac{1}{2} (\Delta_{\psi_{b}} A)^{2}$ respectively) because
they contains no useful information about $\Delta_{\psi_{b}}A$.

Let us now examine a case that is more interesting to us:
 Assume that  $|\psi_{A}\rangle = \delta_{\phi} A |\phi\rangle \neq 0,\;|\psi_{B}\rangle = \delta_{\phi} B|\phi\rangle\neq 0 $,
and   $\delta_{\phi} A |\phi\rangle \,\perp \, \delta_{\phi} B|\phi\rangle $.
Analyzing the right hand sides of inequalities (\ref{n-2t}) and (\ref{n-t4}) with this assumption, one finds that
$\left[\Delta_{\phi}(A + B)\right]^{2} \equiv \left\| \delta_{\phi} A|\phi\rangle + \delta_{\phi} B |\phi\rangle \right\|^{2} \equiv  \left\|\delta_{\phi} A |\phi\rangle\right\|^{2} + \left\|\delta_{\phi} B\phi\rangle\right\|^{2}$. (It is because
$\left\| \,| \psi_{A}\rangle + |\psi_{B}\rangle \right\|^{2} = \left\| \,| \psi_{A}\rangle \right\|^{2} + \left\| \,|\psi_{B}\rangle \right\|^{2}$ for any $|\psi_{A}\rangle\, \perp\,|\psi_{B}\rangle$).
This means that finally, in the case under consideration, the inequality (\ref{n-t4}) takes the following form: $(\Delta_{\phi}A)^{2} + (\Delta_{\phi}B)^{2} \geq \frac{1}{2}\left[(\Delta_{\phi}A)^{2} + (\Delta_{\phi}B)^{2}\right]$ and  we have no information about lower bounds for $(\Delta_{\phi}A)^{2}$ and $(\Delta_{\phi}B)^{2}$. Let us now analyze the inequality (\ref{n-t3}) using the assumption that
$\delta_{\phi} A |\phi\rangle \,\perp \, \delta_{\phi} B|\phi\rangle $. We already know that then $\left[\Delta_{\phi}(A + B)\right]^{2} \equiv (\Delta_{\phi}A)^{2} + (\Delta_{\phi}B)^{2}$. Next,
by multiplying the two sides of the inequality (\ref{n-t3}) by each other respectively, we get that $(\Delta_{\phi}A)^{2} + (\Delta_{\phi}B)^{2} + 2 \Delta_{\phi}A \, \Delta_{\phi}B \geq \left[ \Delta_{\phi}(A + B)\right]^{2} \equiv (\Delta_{\phi}A)^{2} + (\Delta_{\phi}B)^{2} $. This inequality simplifies to $\Delta_{\phi}A \cdot \Delta_{\phi}B \geq 0$, which again gives us no information about the lower bounds on $\Delta_{\phi}A$ and  $ \Delta_{\phi}B$.

In summary, when we study two non--commuting observables, $A$ and $B$, and apply the "sum uncertainty relations" to find lower bounds  on $\Delta_{\phi}A$ and $\Delta_{\phi}B$ in the situations considered above, the conclusions resulting from these relations are the same as the conclusions resulting from the HR uncertainty relations.
The "sum uncertainty relations" seem to be more
useful in the case where we investigate possible relations between lower bounds of the standard deviations, $ (\Delta_{\phi}A_{1}), (\Delta_{\phi}A_{2}), \ldots , (\Delta_{\phi}A_{n})$  of a set of  non--commuting observables $\left\{ A_{j} \right\}_{j=1}^{n}$ with $ n \geq 3$. In such cases, using the generalization of the triangle inequality to $n$ vectors (see, e. g.,  \cite{Cer}),
\begin{equation}
\sum_{j=1}^{n}\left\|\,|\psi_{j}\rangle \right\|\;\geq\; \left\|\sum_{j=1}^{n}|\psi_{j}\rangle \right\|, \label{t-n}
\end{equation}
we obtain the following generalization of the inequality (\ref{n-t2}),
\begin{equation}
\sum_{j=1}^{n}(\Delta_{\phi}A_{j}) \geq   \,\Delta_{\phi}(\sum_{j=1}^{n} A_{j}). \label{t-n1}
\end{equation}
and this type of "sum uncertainty relation" and its various generalizations analogous to (\ref{n-t4}) have recently been intensively studied in many papers (see, e. g. \cite{Pat,Bin,Son} and many others).

\section{Final remarks}

From the analysis carried out in Sec. 2 it follows that the uncertainty relation in its the most general form (\ref{gen-ur}) is in fact an upper bound on the modulus of the correlation function  ${\cal C}_{\phi}(A,B)$.
If $\Delta_{\phi}A > 0$ and  $\Delta_{\phi}B > 0$  then the uncertainty relation (\ref{gen-ur}) can be written  in the equivalent form (\ref{r1}) as the quantum modification of the Pearson correlation coefficient $\mathfrak{r}_{\phi}(A,B)$.
In the literature, the correlation function,   ${\cal C}_{\phi}(A,B)$, or Pearson coefficient, $\mathfrak{r}_{\phi}(A,B)$, are  sometimes used to characterize entanglement of observables $A$ and $B$ in a given state $|\phi\rangle$ of the system (see, e.g. \cite{Khr,Jeb,Mac1,Oh}).
This shows that the uncertainty relation, especially in its most general form  (\ref{gen-ur}) or (\ref{r1}), has a much greater significance in quantum mechanics than its standard understanding as a lower bound on the value of the product of standard deviations $\Delta_{\phi}A\cdot \Delta_{\phi}B$ (or variances).
So in fact the standard Heisenberg--Robertson and Schrodinger uncertainty relations have two faces.
The first is the standard one. According to it, the right-hand side of the inequality (\ref{gen-ur}) is a lower bound of the product $\Delta_{\phi}A\cdot \Delta_{\phi}B$. The non--standard observation is that this lower bound is
the modulus of the correlation function ${\cal C}_{\phi}(A,B)$ of observables $A$ and $B$ in the state $|\phi\rangle$.
Simply, having $A,B$ and $|\phi\rangle$ and using (\ref{co1}) one can calculate the correlation function ${\cal C}_{\phi}(A,B)$ and then, as it follows from (\ref{gen-ur}) one can find
the modulus of this function and, according to the inequality (\ref{gen-ur}), this
modulus will be the lower bound of the product $\Delta_{\phi}A \cdot \Delta_{\phi}B$.
The second  non--standard face is that the minimum of the product $\Delta_{\phi}A \cdot \Delta_{\phi}B$ is an upper bound on the modulus of the correlation function ${\cal C}_{\phi}(A,B)$. Again, for a given  $A, B$ and $|\phi\rangle$,
one can use (\ref{dF}) and calculate standard deviations $\Delta_{\phi}A$ and $ \Delta_{\phi}B$. Then using the inequality (\ref{gen-ur}) one can consider the product $\Delta_{\phi}A \cdot \Delta_{\phi}B$ as the upper
 on the modulus of the correlation function ${\cal C}_{\phi}(A,B)$, which in turn can be used to characterize correlation and entanglement of observables $A$ and $B$ in the state $|\phi\rangle$.
Which of these two possibilities will be used depends on the experiment we want to carry out or the goal we want to achieve.

The correlation function  ${\cal C}_{\phi}(A,B)$ can be equal to zero,  ${\cal C}_{\phi}(A,B) \equiv \langle \phi|\delta_{\phi}A\,\delta_{\phi}B|\phi\rangle = 0$,
when $\delta_{\phi}A|\phi\rangle\neq 0,\; \delta_{\phi}B|\phi\rangle\neq 0$
only if $\dim {\cal H} \geq 3$. This property is impossible to satisfy if $\dim {\cal H} < 3$.
As it also has been shown there may exist  large sets of states $S_{AB} = \{|\phi\rangle \in {\cal H}\,| \,\mathfrak{r}_{\phi}(A,B)=0\} \subset {\cal H}$, (where $\dim {\cal H} \geq 3$),
of a quantum system, which are not eigenstates of any observable from  non--commuting  pairs $A$ and $B$, such that vectors $\delta_{\phi} A |\phi\rangle \neq 0$ and $\delta_{\phi} B|\phi\rangle \neq 0$  are orthogonal: $\delta_{\phi} A |\phi\rangle\,\perp\,\delta_{\phi} B|\phi\rangle $.
These sets may be different for different pairs of such observables.
It should be noted here that
the set $S_{AB}$ is not identical to the set $S_{[A,B]}$ of all vectors for which the right side of the
HR uncertainty
relation (\ref{R1}) is equal to zero: $S_{[A,B]} = \{|\phi\rangle \in {\cal H}\,|\,\mathfrak{r}_{\phi}(A,B) \geq 0\;{\rm and}\; {\cal C}_{\phi}(A,B) =[{\cal C}_{\phi}(A,B)]^{\ast}\}$. Equivalently:
$S_{[A,B]} = \{|\phi\rangle \in {\cal H}\,|\, \langle \phi|[A,B]|\phi\rangle= 0\,{\rm and}\,[A,B] \neq 0,\, \Delta_{\phi}A> 0,\,\Delta_{\phi}B > 0 \}$. There is $S_{AB} \subset S_{[A,B]}$ and  $S_{[A,B]} \setminus S_{AB} \neq \emptyset$ (see Sec. 2).
The set $S_{[A,B]}$ is  a set of states $|\phi\rangle$ for which the correlation function  ${\cal C}_{\phi}(A,B)$ defined   in formula (\ref{co1})  coincides with the classical correlation function.
The state $|\phi_{1}\rangle$ considered in Sec. 3 is an example of a state belonging to the set $S_{AB}$, while the state $|\phi_{2}\rangle$ considered therein  is an example of a vector belonging to the set
$S_{[A,B]}$ that does does not belong to $S_{AB}$.

As can be seen from the above analysis
for states $|\phi\rangle$ belonging to the set $S_{[A,B]}$, the right-hand side of the inequality (\ref{R1}) is equal to zero. However, contrary to popular belief, this does not necessarily mean that the lower bound on the product  $\Delta_{\phi}A \cdot \Delta_{\phi}B$  is equal to zero. The inequality (\ref{gen-ur}) implies that then simply
\begin{equation}
\Delta_{\phi}A \cdot \Delta_{\phi}B \geq |\Re [{\cal C}_{\phi}(A,B)]|, \;|\phi\rangle \in  S_{[A,B]},    \label{gen-ur3}
\end{equation}
and, in general, $\Re [{\cal C}_{\phi}(A,B)]$ needs not be equal to zero if $\Im [{\cal C}_{\phi}(A,B)] =0$, (or, equivalently, if $\langle \phi|[A,B]|\phi\rangle= 0$).  Here, an example is   the above--mentioned state $|\phi_{2}\rangle$.
This means that in such cases the quantum system is viewed by the uncertainty relation HR, (\ref{R1}), just as it would be in an eigenstate of the operator $A$ or $B$,
which can lead to a wrong assessment of the properties of the quantum system in such a state.
The correct description of these properties  can be obtained by using the uncertainty relations (\ref{Sch-2}), (\ref{Sch-3})  proposed by Schrodinger  or the equivalent relation (\ref{gen-ur}).
The HR uncertainty relation (\ref{R1}) gives a correct lower bound for the product $\Delta_{\phi}A \cdot \Delta_{\phi}B$ only for such states $|\phi\rangle$  of the system under studies that  $|\phi\rangle \in S_{\{A,B\}}$, where
$S_{\{A,B\}} = \left\{|\phi\rangle \in {\cal H}| \mathfrak{r}_{\phi}(A,B) \geq 0\;{\rm and}\;\Re [{\cal C}_{\phi}(A,B)] = 0\right\}$.

On vectors belonging to the set $S_{AB}$
 the right--hand sides of the inequalities (\ref{R1}), (\ref{dAdB}) --- (\ref{Sch-3}) reach their absolute minimum, i.e. they have the value zero. So,  the lower bound for the product $\Delta_{\phi}A \cdot \Delta_{\phi}B$ is zero for $|\phi\rangle \in S_{AB}$.
 In other words, quantum theory  (to be more precisely: basic assumptions (postulates) of the quantum theory)  allows
 $\Delta_{\phi}A$ and $\Delta_{\phi}B$
 both to be as small as possible
 and also it allows that in this case the magnitude of $\Delta_{\phi}A $ does not affect the magnitude of  $\Delta_{\phi}B$ and vice versa.
 Non--commuting observables $A$ and $B$ are uncorrelated if the system is in a state $|\phi\rangle \in S_{AB}$ and then they behave as if they were independent:
  In such cases, observable $A$ does not disturb observable $B$ (and vice versa) and does not generate any additional fluctuations of $B$.
This raises the questions:
Can this property be used technically in any way, and if so, how?
What are properties of the system prepared in such a state?

\noindent
{\bf ORCID id:}\\
\noindent
 Krzysztof Urbanowski: https://orcid.org/0000-0002-7490-4278

\section*{Declarations}

\begin{itemize}
\item
This work was supported by
the program of the Polish Ministry
of Science and Higher Education under the name "Regional
Initiative of Excellence", Project No. RID/SP/0050/2024/1.

\item  The author declares no conflict of interest.

\item The author declares that there is not any personal, academic interest, or any
other factors that may be perceived to influence the objectivity, integrity or
value of the study.

\item The author consents to publication.

\item This manuscript has no associated data, or the data will not be deposited.
[Author's comment: This is a theoretical work and analytical calculations
are made. Therefore, no data are required].

\item The author declares that there are no conflicts of interest
regarding the publication of this article  and that all results presented in this article are the author's own results.

\end{itemize}

\end{document}